\documentclass[aps,prd,preprint]{revtex4-1}

\usepackage{epsfig}

\begin{document}

\title{ Effect of gauge boson mass on the phase structure of QED$_{3}$ }
\author{Jian-Feng Li$^{1}$, Yu-Qing Zhou$^{2}$, Hong-Tao Feng$^{2}$, Wei-Min Sun$^{1,3}$ and Hong-Shi Zong$^{1,3}$}
\address{$^{1}$ Department of Physics, Nanjing University, Nanjing 210093, China}
\address{$^{2}$ Department of Physics, Southeast University, Nanjing 211189}
\address{$^{3}$ Joint Center for Particle, Nuclear Physics and Cosmology, Nanjing 210093, China}
\begin{abstract}

Dynamical chiral symmetry breaking (DCSB) in QED$_{3}$ with finite gauge boson mass is
studied in the framework of the rainbow approximation of Dyson-Schwinger equations.
By adopting a simple gauge boson propagator ansatz at finite temperature, we first numerically solve the
Dyson-Schwinger equation for the fermion self-energy to
determine the chiral phase diagram of QED$_3$ with finite gauge boson mass
at finite chemical potential and finite temperature, then we study the
effect of the finite gauge mass on the phase diagram of QED$_3$. It is found
that the gauge boson mass $m_{a}$ suppresses the occurrence of
DCSB. The area of the region in the chiral phase diagram corresponding to DCSB phase decreases as
the gauge boson mass $m_{a}$ increases. In
particular, chiral symmetry gets restored when $m_{a}$ is above a
certain critical value. In this paper, we use DCSB to describe the
antiferromagnetic order and use the gauge boson mass to describe the
superconducting order. Our results give qualitatively a physical
picture on the competition and coexistence between antiferromagnetic
order and superconducting orders in high temperature cuprate superconductors.

\bigskip

\bigskip


\end{abstract}
\maketitle

Quantum electrodynamics in 2+1 dimensions (QED$_3$) has been widely
studied in the last years. It is well known that QED$_{3}$ has two
basic features: dynamical chiral symmetry
breaking (DCSB) in the zero fermion mass limit and confinement \cite{a1,a2,a3,a4,a5,a6,a7,a8}. Moreover, QED$_3$ is
superrenormalizable, so it does not suffer from the ultraviolet
divergence which are present in QED$_4$. Therefore QED$_3$ can be regarded as a useful tool by which one can develop insight into aspects of QCD. In parallel with a toy model of QCD, recently, as a low energy effective theory, QED$_3$ is
also of interest in 2D condensed matter systems, including high-T$_c$ superconductors \cite{a9,a10,a11,a12,a13,a14,a15,a16}, quantum Hall effect \cite{a17} and graphene \cite{a18,a19,a20,a21}.

At zero temperature, by solving a truncated system of Dyson-Schwinger equations, Appelquist et al. first found that DCSB occurs only when the number of fermion flavors $N$ is less than a critical number $N_c$ \cite{a3,a4}.
At finite temperature, Aitchison et al. adopted an approximate treatment of Dyson-Schwinger equation for the the fermion self-energy in the $1/N$ expansion \cite{a22} and found that the critical number of fermion flavors $N_c$ is
temperature-dependent and chiral symmetry is restored above a critical temperature. Recently, Feng et al \cite{a23,a24,b1} investigated the influence of finite chemical potential on the critical number of fermion flavors. It is interesting to study the influence of finite chemical potential and temperature on the critical number of fermion flavor simultaneously. In this paper we shall investigate the phase diagram of QED$_3$ with gauge boson mass $m_{a}$  \cite{a25} at finite chemical potential and temperature. In particular, we study the effect of the gauge boson
mass on phase structure of QED$_3$. In addition, here we use DCSB to describe the
antiferromagnetic (AF) order and use the gauge boson mass to describe
the superconducting (SC) order \cite{a25,b2}, and discuss the competition between
$m_{a}$ and DCSB in chiral phase diagram of QED$_{3}$ with gauge
boson mass. We also note that the phase structure of QED$_{3}$ is also useful for the
study of semimetal-insulator transition in graphene.

In Euclidean space, the Lagrangian of QED$_{3}$ with $N$ massless fermion flavors reads
\begin{equation}\label{eq1}
\mathcal{L}=\sum^N_{i=1}\bar\psi_i(\not\!\partial+\mathrm{i}e\not\!\not\!
A-\gamma_3\mu)\psi_i+\frac{1}{4}F^2_{\rho\nu}+\frac{1}{2\xi}(\partial_\rho\
A_\rho)^2,
\end{equation}
where $\mathcal{L}$ contains the coupling between massless Dirac
fermions and the $U(1)$ gauge field, and $\xi$ is the gauge
parameter. The gauge boson can acquire a finite mass $m_{a}$ via the
Anderson-Higgs mechanism, which changes the original gauge boson
propagator. The $4 \times1$ spinor $\psi_i$ represents the fermion
field, the $4 \times4$ $\gamma_{\mu}$ matrices obey the Clifford
algebra $\{\gamma_\mu,\gamma_\nu\}=2\delta_{\mu\nu}$, and
$i=1,\cdots, N$ are the flavor indices. For a real physical system,
the number of fermion flavors $N$ equals to 2 (the realistic model corresponds to $N=2$, the two spin directions of the electron. see, for instance, \cite{Marston}). So, in this paper we
investigate the chiral phase diagram of QED$_3$ with gauge boson
mass $m_{a}$  at finite chemical potential and temperature
for $N=2$ case. Adopting the rainbow approximation for the
fermion-gauge vertex and neglecting the dependence of the gauge
boson propagator on the chemical potential (this is a commonly used
approximation in studying the dressed fermion propagator at finite
$\mu$ \cite{a23,a24,b1,DSE2,zong2,zong3,DSE5,DSE6,DSE7,DSE8,DSE9}),
we can obtain the Dyson-Schwinger equation for the fermion
propagator at finite chemical potential $\mu$:
\begin{equation}\label{eq2}
S^{-1}(p,\mu)=i\gamma\cdot p-\mu\cdot\gamma_{3}+\frac{\alpha}{N}\int
\frac{d^3k}{(2\pi)^3}\gamma_\rho S(k,\mu)\gamma_\nu D_{\rho\nu}(q),
\end{equation}
where $q=p-k$, $\alpha=e^{2}N$, and $D_{\mu\nu}(q)$ is the gauge boson propagator in Landau gauge
\begin{equation}\label{3}
D_{\mu\nu}(q)=\frac{1}{q^{2}[1+\Pi(q)]+m_{a}^{2}}(\delta_{\mu\nu}-\frac{q_{\mu}q_{\nu}}{q^{2}})
\end{equation}
with $\Pi(q)$ being the vacuum polarization function.

At non-zero chemical potential, according to the treatments in Refs. \cite{a23,a24},
the inverse of the fermion propagator at finite $\mu$ can be obtained from the one at $\mu=0$ by the substitution $p \rightarrow
\tilde{p}=(\vec{p},p_{3}+i\mu)$:
\begin{equation}\label{eq4}
S^{-1}(p,\mu)=S^{-1}(\tilde{p})=i\gamma\cdot
\tilde{p}A(\tilde{p}^{2})+m(\tilde{ p}^{2}),
\end{equation}
where $S^{-1}(p)=i\gamma\cdot
p A(p^{2})+m(p^{2})$ is the inverse of the fermion propagator at $\mu=0$. The validity of the above formula (4) has been discussed in detail in Ref. \cite{CON1}.

Now, let us give a short review of some studies on the effect of the wave function renormalization factor $A(p^2)$. At zero temperature and chemical potential, the results in Eq. (2) show that when the $1/N$ order contribution to $A(p^2)$ is included, the critical fermion flavor number takes almost the same values as the case where $A(p^2)=1$ \cite{a4}. At finite temperature temperature and zero chemical potential, a comparison of studies in Refs. \cite{a22,At} also suggests that the $1/N$ order contribution to $A(p^2)$ only slightly changes the results qualitatively. So, we expect that the the $1/N$ order contribution to $A(p^2)$ is not important at finite temperature and chemical potential and we will take $A(p^2)=1$ in this paper. Then, we can obtain the integral equation for the dynamically generated mass function:
\begin{equation}\label{eq5}
m(\tilde{p}^{2})
=\frac{\alpha}{4N}\int\frac{d^{3}k}{(2\pi)^{3}}Tr[\gamma_{\rho}\frac{-i\gamma\cdot\tilde{k}+m(\tilde{k}^{2})}{\tilde{k}^{2}+m(\tilde{k}^{2})}\gamma_{\nu}D_{\rho\nu}(q)].
\end{equation}

At non-zero temperature, for the purpose of obtaining a qualitative picture of dynamical picture of dynamical mass generation, we employ the following simplified gauge boson propagator ansatz \cite{a26,a27} in the presence of a finite gauge boson mass
\begin{equation}\label{eq6}
\Delta_{\mu\nu}(q_0,Q,\beta)=\frac{\delta_{\mu3}\delta_{\nu3}}{Q^{2}+\Pi_{0}(Q,\beta)+m_{a}^{2}},
\end{equation}
where $\Pi_{0}(Q,\beta)$ is defined as:
\begin{equation}\label{eq7}
\Pi_{0}(Q,\beta)=\frac{\alpha}{8\beta}[Q\beta+\frac{16\ln2}{\pi}\exp(-\frac{\pi\beta}{16\ln2})].
\end{equation}
At finite temperature, the integration over the third component of a fermion loop momentum is replaced by an infinite sum over odd Matsubara frequencies. So the integral equation for the dynamically generated mass function at finite temperature and chemical potential is given by
\begin{equation}\label{eq8}
m(\beta,\mu,P)=\frac{\alpha}{N\beta}\sum_{n=-\infty}^{\infty}\int\frac{d^{2}k}{(2\pi)^{2}}\frac{1}{Q^{2}+\Pi_{0}(\beta,Q)+m_{a}^{2}}\frac{m(\beta,\mu,K)}{[K^{2}+m^{2}(\beta,\mu,K)+(\varpi_{n}+i\mu)^{2}]},
\end{equation}
where
\begin{eqnarray}\label{eq9}
p=(p_{0},~\textbf{p}),~P=|\textbf{p}|,~p_{0}=(2m+1)\frac{\pi}{\beta},\nonumber\\
k=(k_{0},~\textbf{k}),~K=|\textbf{k}|,~k_{0}=(2n+1)\frac{\pi}{\beta},\nonumber\\
q=(q_{0},~\textbf{q}),~Q=|\textbf{q}|=|\textbf{p}-\textbf{k}|,q_{0}=2(m-n)\frac{\pi}{\beta}.
\end{eqnarray}
The sum over infinite Matsubara frequencies can be done analytically:
\begin{eqnarray}\label{eq13}
&&\sum_{n=-\infty}^{\infty}\frac{1}{(\varpi_{n}+i\mu)^{2}+K^{2}+m^{2}(\beta,\mu,K)}\\
&&=\frac{1}{2\sqrt{K^{2}+m^{2}(\beta,\mu,K)}}\times
\sum_{n=-\infty}^{\infty}[\frac{1}{i\varpi_{n}+\sqrt{K^{2}+m^{2}(\beta,\mu,K)}}-\frac{1}{i\varpi_{n}-\sqrt{K^{2}+m^{2}(\beta,\mu,K)}}]\nonumber\\
&&=\frac{\beta}{2\sqrt{K^{2}+m^{2}(\beta,\mu,K)}}[\frac{1}{e^\beta(\mu-\sqrt{K^{2}+m^{2}(\beta,\mu,K)})+1}-\frac{1}{e^\beta(\mu+\sqrt{K^{2}+m^{2}(\beta,\mu,K)})+1}],\nonumber
\end{eqnarray}
where we have made use of the formula
\begin{equation}
\sum_{n=-\infty}^{\infty}\frac{1}{i\varpi_{n}-x}=\frac{\beta}{e^{\beta x} +1}.
\end{equation}
Substituting Eq. (10) into Eq. (8), we can obtain the integral equation for the mass function with finite gauge boson mass at finite $T$ and $\mu$
\begin{eqnarray}\label{eq14}
m(\beta,\mu,P)=\frac{\alpha}{8 N\pi^{2}}\int
d^{2}K\frac{m(\beta,\mu,K)}{\sqrt{K^{2}+m^{2}(\beta,\mu,K)}\times
[Q^{2}+\Pi_{0}(\beta,Q)+m_{a}^{2}]} \nonumber\\
 \times
[\frac{1}{e^\beta(\mu-\sqrt{K^{2}+m^{2}(\beta,\mu,K)})+1}-\frac{1}{e^\beta(\mu+\sqrt{K^{2}+m^{2}(\beta,\mu,K)})+1}].
\end{eqnarray}
Here we also note that in graphene system the gap equation has a similar form \cite{a21}:
\begin{eqnarray}\label{eq15}
m(\beta,\mu,P)&&=\frac{\alpha}{8 N\pi^{2}}\int
d^{2}K\frac{m(\beta,\mu,K)\times V(Q,\mu)}{\sqrt{K^{2}+m^{2}(\beta,\mu,K)}}\nonumber\\
&&\times[\frac{1}{e^\beta(\mu-\sqrt{K^{2}+m^{2}(\beta,\mu,K)})+1}-\frac{1}{e^\beta(\mu+\sqrt{K^{2}+m^{2}(\beta,\mu,K)})+1}],
\end{eqnarray}
where the Coulomb interaction
$V(Q,\mu)=\frac{1}{[Q/8\lambda+\Pi(\mu,Q)/N]}$ with the coupling
strength $\lambda$ and the polarization function being given in Ref. \cite{a21}. In the context of graphene, the dominant interaction for Dirac fermions is the long-range Coulomb interaction, rather than gauge interaction. At finite chemical potential $\mu$, however, the polarization function is proportional to $\mu$ as $q$ tends to zero, which implies that the Coulomb interaction between Dirac fermions is statically screened and thus becomes short-ranged. The static screening factor $\mu$ appearing in the Coulomb interaction function in graphene system plays a similar role as the finite gauge boson mass $m_a$ in the effective QED$_{3}$ theory of high-$T_{c}$ superconductor. For example, $\mu$
suppresses semimetal-insulator transition in graphene system, while $m_a$ suppressed DCSB in QED$_{3}$ theory of high-$T_{c}$
superconductor.

\begin{figure}[htp!]
\includegraphics[width=10cm]{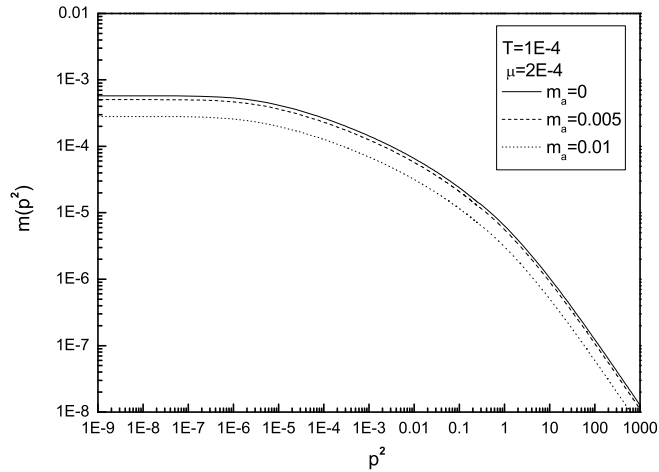}
\vspace{-0.8cm}
\caption{$m(p^2)$ for several values of the gauge boson mass $m_{a}$ at $N=2$, $T=0.0001$ and $\mu=0.0002$.}\label{FIG}
\end{figure}

Now we employ the iteration algorithm to numerically solve Eq. (12). In the numerical calculation we set $\alpha= 1$, thus every quantity with mass dimension is scaled and becomes dimensionless. In the real numerical calculation, we adopt an ultraviolet cutoff $\Lambda=10^{5}$, which is large enough to ensure that the calculated results are stable with respect to $\Lambda$.
The calculated dynamically generated mass function $m(p^2)$ is shown in Fig. 1 for $N=2$, $T=0.0001$ and $\mu=0.0002$, and for several values of the
gauge boson mass. From Fig. 1 it can be seen that the dynamically generated mass $m(p^{2})$ is almost constant for
small $p^2$ and tends to zero at large $p^2$. This result is consistent with that given in Ref. \cite{a28}. In addition, we notice that $m(p^2)$ decreases with the gauge boson mass increasing for any fixed value of $p^2$. This reflects the following fact: Once the gauge boson acquires a finite mass via Anderson-Higgs mechanism , it cannot mediate a
long-range interaction and suppresses the condensation of fermion-antifermion pairs.

The phase diagram of QED$_{3}$ obtained in the presence of a small gauge boson mass for $N = 2$ is plotted in Fig. 2. It is clearly seen that for each fixed value of the gauge boson mass the phase curve separates the DCSB phase (Nambu
phase, where $m(p^{2})> 0$) and the chiral symmetric phase (Wigner phase, where $m(p^{2}) = 0$). It can also be seen that the area of the region corresponding to the DCSB phase becomes small with $m_{a}$ increasing. This is understandable, because a finite gauge boson mass $m_{a}$ weakens the gauge interaction and suppresses the occurrence of DCSB.

\begin{figure}[htp!]
\includegraphics[width=10cm]{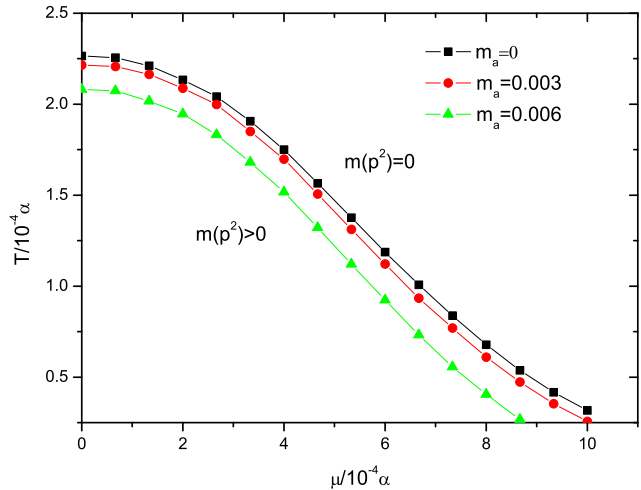}
\vspace{-0.8cm}
\caption{The phase diagram of QED$_{3}$ for several gauge boson
 mass $m_{a}$}\label{FIG}
\end{figure}

From Fig. 2, it can be seen that when $m_{a}$ is above a critical value $m_{a}^{crit}\approx 0.018$,
the DCSB phase will rapidly disappear on the $\mu-T$ plane. This
result gives a simple physical picture on the competition between
the AF order and the SC order in high temperature cuprate
superconductors, where the AF order dominates at low doping while
the SC order dominates at higher doping. In addition, from Fig. 2 it is easy to find
that DCSB disappears and chiral symmetry gets restored when $\mu$
(or $T$) is above a certain critical value $\mu^{crit}$ (or
$T^{crit}$), and the critical temperature decreases
as the chemical potential $\mu$ increases.

Just as was shown above, in QED$_3$ it is the dimensional coupling constant $\alpha$ (which is set to be 1 in our calculation) that sets the energy scale of the system. From our numerical results, it can be seen that the dynamically generated fermion mass, the critical temperature and chemical potential are quite small compared with this energy scale. 
Then, what is the reason for this? As far as we know, there is no simple way of seeing why the dynamically generated mass, the critical temperature and chemical potential are substantially less than the natural scale $\alpha$. However, this is what seems to happen in all such calculations. For example, in the BCS calculation of the superconducting energy gap at $T=0$, it is found that the zero-temperature gap, $\Delta$, is given by $\Delta \approx 2 \omega_D e^{-\frac{1}{V N_F}}$, where $\omega_D$ is the Debye (cut-off) frequency, $V$ is the volume and $N_F$ is the density of states. The quantity $V N_F$ plays the role of a dimensionless interaction parameter, and enters in a non-analytic way in the expression for $\Delta$. This is essentially because the calculation is (strongly) non-perturbative. The natural scale $\omega_D$ is substantially modified by the non-perturbative exponential factor (for details, see, for instance, Ref. \cite{Aitchison}). This exponential factor multiplying the natural scale is very similar to what happens in the $T=0$ QED$_3$ calculation. Something similar seems to be true at finite $T$ as well.
The fact that such calculations lead to dynamically generated symmetry breaking masses which are substantially smaller than a natural energy scale might help to explain the "hierarchy problem", namely the wide discrepancy between (say) Grand Unification scales and the weak scale (for instance, the idea that QED$_3$ might be a "laboratory" for
studying the origin of different mass scales through non-perturbative dynamics was suggested in Ref. \cite{Pennington}).

It is well known that the cuprate superconductor is a Mott insulator
with long-range AF order at half filling. Once
holes are doped into the CuO$_{2}$ planes, SC order occurs and the long-range AF order
disappears in the material. Recently, elaborate neutron scattering and scanning
tunneling microscopy experiments \cite{a29,a30} have found the
competition between AF correlation and SC order in high-$T_{c}$
cuprate superconductor. It is clear that the gauge boson will become
massive once SC appears in the cuprate superconductor. Here, in order to
show the competition between the AF order and the SC order, following
the approach of Ref. \cite{b2}, we use the gauge boson mass
to describe the SC order and use DCSB to describe the AF order.

To summarize, in this paper, in the framework of the rainbow approximation of the Dyson-Schwinger equation and employing the approximation of ignoring the $\mu$ dependence of the photon propagator and setting the wave function renormalization factor $A(p^{2})=1$ , we obtain the gap equation of QED$_3$ at finite temperature and chemical potential. We
investigate the phase diagram of QED$_3$ for several values of gauge boson mass $m_{a}$ at finite chemical potential and temperature for $N=2$ case, which corresponds to real physical systems. In particular, we study the effect of a finite gauge boson
mass on phase structure of QED$_3$ and find that a finite gauge boss mass
$m_{a}$ suppresses the DCSB, which shows a competition between antiferromagnetic order and superconducting orders in high
temperature superconductors. Besides, by a comparison between Eq. (14) and Eq. (15), we expect that the excitionic semimetal-insulator transition in graphene can be well described by the phase structure of QED$_{3}$.

\begin{acknowledgements}
We are grateful to G.Z. Liu and I.J.R. Aitchison for helpful discussions. This work is supported in part by the National Natural Science Foundation of China (under Grant Nos. 10775069 and 10935001) and the Research Fund for the
Doctoral Program of Higher Education (Grant No. 200802840009).
\end{acknowledgements}


\begin{references}
\bibitem{a1}  J.M. Cornwall, Phys. Rev. \textbf{D 22}, 1452 (1980).
\bibitem{a2}  R.D. Pisaski,  Phys. Rev. \textbf{D 29}, 2423 (1984).
\bibitem{a3}  T. Appelquist, M. Bowick, D. Karabali, and L.C.R. Wijewardhana, Phys. Rev. \textbf{D 33}, 3704 (1986).
\bibitem{a4}  T. Appelquist, D. Nash, and L.C.R. Wijewardhana, Phys. Rev. Lett. \textbf{60}, 2575 (1988).
\bibitem{a5}  D. Nash, Phys. Rev. Lett. \textbf{62}, 3024 (1989).
\bibitem{a6}  P. Maris, Phys. Rev. \textbf{D 52}, 6087 (1995).
\bibitem{a7}  P. Maris, Phys. Rev. \textbf{D 54}, 4049 (1996).
\bibitem{a8}  A. Bashir, A. Raya, I.C. Clo$\ddot{e}$t, and C.D. Roberts, Phys. Rev. \textbf{C 78}, 055201 (2008).
\bibitem{a9}  P.A. Lee, N. Nagaosa, and C.G. Wen, Rev. Mod. Phys. \textbf{78}, 17 (2006).
\bibitem{a10} W. Rantner and X.G. Wen, Phys. Rev. Lett. \text{69}, 1811 (1992).
\bibitem{a11} D.H. Kim, P.A. Lee and X.G. Wen, Phys. Rev. Lett. \textbf{76}, 503 (1996).
\bibitem{a12} D.H. Kim and  P.A. Lee, Ann. Phys. \textbf{272}, 130 (1999).
\bibitem{a13} M. Franz and Z. TesAanovic, Phys. Rev. Lett. \textbf{87}, 257003 (2001).
\bibitem{a14} I.F. Herbut, Phys. Rev. Lett. \textbf{88}, 047006 (2002).
\bibitem{a15} I.F. Herbut, Phys. Rev. \textbf{B 66}, 094504 (2002).
\bibitem{a16} D.J. Lee and I.F. Herbut, Phys. Rev. \textbf{B 66}, 094512 (2005).
\bibitem{a17} X.G. Wen and A. Zee, Phys. Rev. Lett. \textbf{86}, 3871 (2001).
\bibitem{a18} V.P. Gusynin, S.G. Sharapov et al, Phys. Rev. lett. \textbf{95}, 146801 (2005).
\bibitem{a19} V.P. Gusynin, S.G. Sharapov et al, Phys. Rev. lett. \textbf{96}, 256802 (2006).
\bibitem{a20} D.V. Khveshchenko, Phys. Rev. lett. \textbf{87}, 246802 (2001).
\bibitem{a21} G.Z. Liu, W. Li, and G. Cheng, Phys. Rev. \textbf{B 79}, 205429 (2009).
\bibitem{a22} I.J.R. Aitchison, N. Dorey et al, Phys. Lett. \textbf{B 294}, 91 (1992).
\bibitem{a23} H.T. Feng, F.Y. Hou, X. He, W.M. Sun, and H.S. Zong, Phys. Rev. \textbf{D 73}, 016004 (2006).
\bibitem{a24} H.T. Feng, W.M. Sun, D.K. He, and H.S. Zong, Phys. Lett. \textbf{B 661}, 57 (2008).
\bibitem{b1}  H.T. Feng, M. He, W.M. Sun, and H.S. Zong,  Phys. Lett. {\bf B 688}, 178 (2010).
\bibitem{a25} T.P. Barnea and M.Franz, Phys. Rev. \textbf{B 67}, 060503 (2003).
\bibitem{b2}  G.Z. Liu and G. Cheng, Phys. Rev. \textbf{D 67}, 065010 (2003); G.Z. liu and G. Cheng, arXiv:cond-mat/0208061
\bibitem{Marston} J.B. Marston and I. Affleck, Phys. Rev. \textbf{B 39}, 11538 (1989).
\bibitem{a28} M. He, H.T .Feng, W.M. Sun, and H.S. Zong, Mod. Phys. Lett. \textbf{A 22}, 449 (2007).
\bibitem{DSE2} C.D. Roberts and S.M. Schmidt, Prog. Part. Nucl. Phys. \textbf{45S1}, 1 (2000), and references therein.
\bibitem{zong2}H.S. Zong, L. Chang, F.Y. Hou, W.M. Sun, and Y.X. Liu, Phys. Rev. C {\bf 71}, 015205 (2005).
\bibitem{zong3}F.Y. Hou, L. Chang, W.M. Sun, H.S. Zong, and Y.X. Liu, Phys. Rev. C {\bf 72}, 034901 (2005).
\bibitem{DSE5} Y. Taniguchi and Y. Yoshida, Phys. Rev. D {\bf 55}, 2283 (1997).
\bibitem{DSE6} D. Blaschke, C.D. Roberts, and S. Schmidt, Phys. Lett. B {\bf 425}, 232 (1998).
\bibitem{DSE7} P. Maris, C.D. Roberts, and P.C. Tandy, Phys. Lett. B {\bf 420}, 267 (1998).
\bibitem{DSE8} A. Bender, W. Detmold, and A.W. Thomas, Phys. Lett. B {\bf 516}, 54 (2001).
\bibitem{DSE9} O. Miyamura, S. Choe, Y. Liu, T. Takaishi, and A. Nakamura, Phys. Rev. D {\bf 66}, 077502 (2002).
\bibitem{CON1} Y. Jiang, Y.M. Shi, H.T. Feng, W.M. Sun, and H.S. Zong, Phys. Rev. C {\bf 78}, 025214 (2008).
\bibitem{At}   I.J.R. Aitchison and M. Klein-Kreisler, Phys. Rev. \textbf{D 50}, 1068 (1994).
\bibitem{a26}  N. Dorey and N.E. Mavromatos, Nucl. Phys. \textbf{B 386}, 614 (1992).
\bibitem{a27}  N. Dorey and N.E. Mavromatos, Phys. Lett. \textbf{B 266}, 163 (1991).
\bibitem{Aitchison} I.J.R. Aitchison and A.J.G. Hey, {\it Gauge Theories in Particle Physics}, Vol. 2 (Bristol and Philadelphia: IoP Publishing, 2004), section 17.7.
\bibitem{Pennington} M.R. Pennington and D. Walsh, Phys. Lett. B {\bf 253}, 246 (1991).
\bibitem{a29}  B. Lake et al., Science. \textbf{291}, 1759 (2001); Nature (London). \textbf{415}, 299 (2002).
\bibitem{a30}  J.E. Hoffman et al., Science \textbf{295}, 466 (2002).

\end{references}
\end{document}